\documentclass[reprint,nofootinbib,superscriptaddress,amsmath,amssymb,aps]{revtex4-1}
\usepackage{graphicx}
\usepackage{rotating}
\usepackage{dcolumn}
\usepackage{bm}
\usepackage{color}
\usepackage{mathptmx, textcomp}
\usepackage[latin1]{inputenc}
\usepackage{braket}
\usepackage[colorlinks,linkcolor=magenta,anchorcolor=cyan,citecolor=blue]{hyperref}
\usepackage[multiple]{footmisc}

\def\be{\begin{equation}}
\def\ee{\end{equation}}
\def\ba{\begin{eqnarray}}
\def\ea{\end{eqnarray}}

\def\nn{\nonumber}
\def\lf{\left}
\def\rt{\right}

\newcommand{\eq}[1]{(\ref{#1})}

\def\lf{\left}\def\rt{\right}\def\q{\theta} \def\w{\omega}     \def\p {\pi} \def\a {\alpha}  \def\d {\delta} \def\f {\phi}    \def\k {\kappa} \def\l {\lambda} \def\z {\zeta} \def\x {\xi} \def\c {\chi} \def\b {\beta}  \def\m {\mu} \def\pd {\partial}\def\p {\pi} \def \inf {\infty}  
\def\Q{\Theta} \def\W{\Omega}     \def\S {\Sigma}  \def\F {\Phi} \def\G {\Gamma}     \def\grad{\nabla}\def\.{\cdot}
\def\math {\mathcal}
\begin{document}

\title{Static charged Gauss-Bonnet black holes cannot be overcharged by the new version of gedanken experiments}
\author{Jie Jiang}
\email{jiejiang@mail.bnu.edu.cn}
\affiliation{Department of Physics, Beijing Normal University, Beijing, 100875, China}

\date{\today}

\begin{abstract}
Based on the new version of the gedanken experiments proposed by Sorce and Wald, we examine the weak cosmic censorship conjecture (WCCC) under the spherically charged infalling matter collision process in the static charged Gauss-Bonnet black holes. After considering the null energy condition and assuming the stability condition, we derive the perturbation inequality of the matter source. As a result, we find that the static charged Gauss-Bonnet black holes cannot be overcharged under the second-order approximation of the perturbation when the null energy condition is taken into account, although they can be destroyed in the old version of gedanken experiments. Our result shows that the WCCC holds for the above collision process in the Einstein-Maxwell-Gauss-Bonnet gravity and indicates that WCCC may also be valid in the higher curvature gravitational theories.

\end{abstract}
\maketitle

\section{Introduction}

The weak cosmic censorship conjecture (WCCC) is one of the most important open questions in classical gravitational theory. It states that the singularities must be surrounded by the event horizon and cannot be seen by distant observes \cite{RPenrose}. One way of examining this conjecture is to see whether the event horizon can be destroyed by some physical processes. In 1974, Wald suggested a gedanken experiment and proved that the extremal Kerr-Newman black hole cannot be destroyed via dropping a test particle \cite{Wald94}. However, there are two crucial assumptions in his setup: the background black hole is extremal and the analysis is only at the level of the first-order perturbation. Therefore, Hubeny \cite{Hubeny} illustrated that the near extremal Kerr-Newman black holes can be destroyed under the second-order approximation of the test particle. This attracted lots of attention and was then studies in various theories\cite{B1, B2, B3, B4, B5, B6,B7,B8,B9,B10,B11,B12,B13,B14,B15,B16,B17}.

However, the method which performs a test particle has some inherent defects. In the process of the particle dropping into the black hole, the spacetime is just treated as a background. When considering the second-order approximation, the backreaction, self-force and finite size effects should be taken into consideration. To solve the above defects, Sorce and Wald \cite{SW} has proposed a new version of the gedanken experiments to examine the WCCC, in which they consider a full dynamical process of the spacetime and collision matter fields based on the Iyer-Wald formalism \cite{IW}. After the null energy condition of the collision matter fields is taken into account, they derived a second-order perturbation inequality on the second-order correction of black hole mass $\d^2 M$. As a result, they showed that the nearly extremal Kerr-Newman black hole cannot be destroyed under the second-order approximation after considering this new inequality. Moreover, because the process considered is full dynamical, all self-force, finite size effects, and backreaction are automatically taken into account \cite{IW}.

Most recently, following the setup of this new version, WCCC has also been investigated in the five-dimensional Myers-Perry black holes \cite{An:2017phb}, higher-dimensional charged black holes \cite{Ge:2017vun}, charged dilaton black holes \cite{Jiang:2019ige}, RN-AdS black holes \cite{WJ}, static Einstein-Born-Infeld black hole, Kerr-Sen black holes \cite{Jiang:2019vww} as well as the scalar-hairy RN black holes \cite{Jiang:2020btc}. Although all of them showed the validity of the WCCC in the new version of gedanken experiments, there is still a lack of the general proof of the WCCC. Therefore, it is necessary for us to test it in various theories. We can see that all of the analyses above are performed in the context of the Einstein gravity, relatively little is in higher curvature gravitational theories. After the quantum effect or string modification are taken into account, higher curvature term should be added to the Einstein-Hilbert action. In order for this story to be truly consistent, it is necessary to check whether the WCCC is also valid in higher curvature gravity. As one of the most interest higher curvature gravitational theories, the Einstein-Maxwell-Gauss-Bonnet (EMGB) gravity is an important generalization of Einstein gravity, where the Gauss-Bonnet term can be regarded as a correction from the heterotic string theory \cite{DWitten,Zumino}. In \cite{Ghosh:2019dzq}, the authors investigated the old version of the gedanken experiments in the charged static Gauss-Bonnet black holes and found that the near extremal case can also be overcharged. Actually, they also neglected the self-force, finite size, and backreaction effects in their discussion. Therefore, in this paper, we would like to check whether the WCCC can be restored in the EMGB gravity after the second-order perturbation inequality is considered.

Our paper is organized as follows. In the next section, we
discuss the spacetime geometry of static charged Gauss-Bonnet black holes perturbed by the spherically infalling matter source. In Sec. \ref{sec3}, based on the Iyer-Wald formalism as well as the null energy condition, we derived the perturbation inequality of the matter fields under the second-order approximation. In Sec. \ref{sec4}, we utilize the new version of the gedanken experiment to destroy the near extremal static charged Gauss-Bonnet black holes under the second-order approximation of perturbation. Finally, the conclusions are presented in Sec. \ref{sec5}.

\newpage

\section{Perturbed charged static Gauss-Bonnet black hole geometry}\label{sec2}

In this paper, we would like to investigate WCCC for $n$-dimensional charged Gauss-Bonnet black holes in Einstein-Maxwell-Gauss-Bonnet (EMGB) gravity. The action of the EMGB gravity can be expressed as \cite{GB1,GB2,GB3}
\ba\begin{aligned}
\bm{L}=\frac{\bm{\epsilon}}{16\p}\lf[R+\a_\text{GB} \math{L}_\text{GB}-F_{ab}F^{ab}\rt]+\bm{L}_\text{mt}\,.
\end{aligned}\ea
with
\ba\begin{aligned}
\math{L}_\text{GB}=R^2-4R_{ab}R^{ab}+R_{abcd}R^{abcd}\,,
\end{aligned}\ea
in which $\bm{F}=d\bm{A}$ is the strength of the electromagnetic field $\bm{A}$, $\a_\text{GB}$ is the Gauss-Bonnet coupling constant, $\bm{\epsilon}$ is the volume element, and $\bm{L}_\text{mt}$ is the Lagrangian of the extra matter fields. The equation of motion (EOM) of the EMGB gravity is given by
\ba\begin{aligned}
G_{ab}-\frac{\a_\text{GB}}{2}H_{ab}&=8\p \lf(T_{ab}^\text{EM}+T_{ab}\rt)\,,\\
\grad_a F^{ab}&=4\p j^b\,,
\end{aligned}\ea
with
\ba\begin{aligned}
H_{ab}&=g_{ab}\math{L}_\text{GB}-4R R_{ab}+8R_{ac}R^c{}_b\\
&+8R_{acbd}R^{cd}-4R_{acde}R_b{}^{cde}\,,\\
T_{ab}^\text{EM}&=\frac{1}{4\p}\lf[F_{ac}F_b{}^c-\frac{1}{4}g_{ab}F_{cd}F^{cd}\rt]\,.
\end{aligned}\ea
Here $G_{ab}$ is the Einstein tensor, $T^\text{EM}_{ab}$ and $T_{ab}$ are the stress-energy tensor of the electromagnetic field and perturbation matter fields, separately, and $j^a$ is electric current of the perturbation matter fields. The static solution of above theory can be written as  \cite{GB1,GB2,GB3}
\ba\begin{aligned}
ds^2&=-f(r)dv^2+2dr dv+r^2d\W^2_{n-2}\,,\\
\bm{A}&=\lf(\F_0-\frac{4\p Q}{(n-3)\W r^{n-3}}\rt)dv\,
\end{aligned}\ea
with the blackening factor
\ba\begin{aligned}
f(r)&=1+\frac{r^2}{2\a}\lf(1-\sqrt{1+\frac{4\a \m}{r^{n-1}}-\frac{4\a q^2}{r^{2n-4}}}\rt)\,,
\end{aligned}\ea
and the line element of the unit $(n-2)$-dimensional sphere
\ba\begin{aligned}
d\W_{n-2}^2=\sum_{i=1}^{n-2}\left[\left(\prod_{j=1}^{i-1}\sin^2\q_j\right)d\q_i^2\right]\,.
\end{aligned}\ea
Here the constant $\a$ is related the Gauss-Bonnet coupling constant as $\a=(n-3)(n-4)\a_\text{GB}$, $\W=2\p^{(n-1)/2}/\G[(n-1)/2]$ denotes the volume of the unit $(n-2)$-dimensional sphere, and $\F_0$ is a constant related the gauge freedom of the electrodynamics. The parameters $\m$ and $q$ are the parameters related to the mass and charge of the black hole as
\ba\begin{aligned}
\m=\frac{16\p M}{(n-2)\W}\,,\ \ \ q=\frac{4\p}{\W}\sqrt{\frac{2}{(n-2)(n-3)}}Q\,,
\end{aligned}\ea
The horizon of the black hole is determined by the equation $f(r)=0$ and the radius of the event horizon is the largest root of blackening factor $f(r)$. The surface gravity, area, and electric potential of the static charged Gauss-Bonnet black hole are given by
\ba\begin{aligned}
\k=\frac{f'(r_h)}{2}\,,\ \ \ A_H=\W r_h^{n-2}\,,\ \ \ \F_H=\frac{4\p Q}{(n-3)\W r_h^{n-3}}\,.
\end{aligned}\ea

If the blackening factor satisfies the condition $f'(r_h)>0$, there exists two Killing horizons of the black hole solutions. However, if $f'(r_h)=0$, these two horizons overlap and the black hole becomes extreme. For the extremal case, the conserved quantities of the black hole satisfy the constraints $f(r_e)=f'(r_e)=0$ with the horizon radius $r_h=r_e$, and the constraints can be explicitly express as
\ba\begin{aligned}
\m&=2r_e^{n-5}\lf(r_e^2+\frac{(n-4)\a}{n-3}\rt)\,,\\
q^2&=r_e^{2(n-4)}\lf(r_e^2+\frac{(n-5)\a}{n-3}\rt)\,.
\end{aligned}\ea

In the following, we consider a one-parameter family spherically charged infalling matter perturbation in the Gauss-Bonnet black hole, where the perturbation matter fields are only non-zero in a compact region of the future horizon and vanishing on the bifurcation surface $B$. Under this perturbation, the dynamical fields $\f(\l)$ are labeled by the variation parameter $\l$. Here we denote $\f$ to $g_{ab}$, $\bm{A}$ and the other fields related to the matter source. Then, the EOM in this setup can be expressed as
\ba\begin{aligned}
G_{ab}(\l)-\frac{\a}{2}H_{ab}(\l)&=8\p \lf[T_{ab}^\text{EM}(\l)+T_{ab}(\l)\rt]\,,\\
\grad_a^{(\l)} F^{ab}(\l)&=4\p j^b(\l)\,,
\end{aligned}\ea
where $\grad_a^{(\l)}$ denotes the derivative operator related to the metric $g_{ab}(\l)$, $T_{ab}(\l)$ and $j^b(\l)$ are only nonvanishing in a compact region. Since the case we considered here only contains an spherically symmetric infalling matter source, the spacetime can be generally described by the Vaidya solution with the line element
\ba\begin{aligned}\label{vds}
ds^2(\l)=-f^{(\l)}(r,v) dv^2+2dv dr+r^2d\W^2_{n-2}\,.
\end{aligned}\ea
Because the background spacetime is Gauss-Bonnet black hole, we have $f^{(0)}(r,v)=f(r)$ and $T_{ab}(0)=j^a(0)=0$. For simplification, in the following, we will denote $\c=\c(0)$ to the quantity $\c$ in the background, and its first two order variation is expressed by
\ba\begin{aligned}
\d\c=\left.\frac{d\c}{d\l}\right|_{\l=0}\,,\ \ \ \d^2\c=\left.\frac{d^2\c}{d\l^2}\right|_{\l=0}\,.
\end{aligned}\ea

With similar consideration of Ref. \cite{SW}, here we also assume that the static charged Gauss-Bonnet black hole is linearly stable to above perturbation, which means that after a sufficient time of the perturbation, the spacetime can also be described by the static charged Gauss-Bonnet solution. Then, we choose a hypersurface $\S=\S_0\cup \S_1$. Here $\S_0$ is a portion of the future horizon in the background spacetime (i.e., the hypersurface with radius $r=r_h$ for the background spacetime horizon $r_h=r_h(0)$), and it is bounded by the bifurcation surface $B$ as well as the very late cross section $B_1$. $\S_1$ is a spacelike hypersurface connected $B_1$ and spatial infinity where the dynamical fields are described by static charged Gauss-Bonnet solution. Note that $\S_0$ (i.e., $r=r_h$) is not the horizon of the spacetime with the metric $g_{ab}(\l)$. For later convenience, by considering the gauge freedom of the electromagnetic field, we impose the gauge condition such that
\ba\begin{aligned}
\left.\x^a A_a(\l)\right|_{r=r_h}=0\,,
\end{aligned}\ea
with the Killing vector $\x^a=(\pd/\pd v)^a$ of the background spacetime. Then, under this gauge condition, the dynamical fields can be expressed as
\ba\begin{aligned}\label{dsAS}
ds^2(\l)&=-f^{(\l)}(r)dv^2+2dr dv+r^2d\W^2_{n-2}\,,\\
\bm{A}&=\frac{4\p Q(\l)}{(n-3)\W}\lf(\frac{1}{r_h^{n-3}}-\frac{1}{r^{n-3}}\rt)dv\,
\end{aligned}\ea
with the blackening factor
\ba\begin{aligned}
f^{(\l)}(r)&=1+\frac{r^2}{2\a}\lf(1-\sqrt{1+\frac{4\a \m(\l)}{r^{n-1}}-\frac{4\a q^2(\l)}{r^{2n-4}}}\rt)\,.
\end{aligned}\ea
on the hypersurface $\S_1$. The strength of the electromagnetic field on $\S_1$ is given by
\ba\begin{aligned}
\bm{F}(\l)=\frac{4\p Q(\l)}{\W r^{n-2}}dr\wedge dv\,.
\end{aligned}\ea

\section{Perturbation inequality}\label{sec3}

In this section, we would like to derive a perturbation inequality of above collision matter sources. Following the calculation as \cite{SW}, we also consider the off-shell variation of the EMGB gravity neglected the matter fields part, i.e.,
\ba\begin{aligned}\label{action}
\bm{L}=\frac{\bm{\epsilon}}{16\p}\lf(R+\a_\text{GB} \math{L}_\text{GB}-F_{ab}F^{ab}\rt)\,.
\end{aligned}\ea
The variation of above action gives
\ba\begin{aligned}\label{var1}
\d \bm{L}=\bm{E}_\f \d\f+d\bm{\Q}(\f,\d\f)\,,
\end{aligned}\ea
where
\ba\begin{aligned}
\bm{E}_\f\d\f&=-\bm{\epsilon}\lf(\frac{1}{2}T^{ab}\d g_{ab}+j^a\d A_a\right)\,,\\
\bm{\Q}(\f,\d\f)&=\bm{\Q}^\text{GB}(\f,\d\f)+\bm{\Q}^\text{EM}(\f,\d\f)
\end{aligned}\ea
with the symplectic potential of the gravity part and electromagnetic part
\ba\begin{aligned}\label{Q2}
\bm{\Q}_{a_1\cdots a_{n-1}}^\text{GB}&=\frac{1}{8\p}\bm{\epsilon}_{ba_1\cdots a_{n-1}}\lf(P_a{}^{cbd}\d \G^a{}_{cd}+\d g_{bd}\grad_a P^{acbd}\rt)\,,\\
\bm{\Q}_{a_1\cdots a_{n-1}}^\text{EM}&=-\frac{1}{4\p}\bm{\epsilon}_{ba_1\cdots a_{n-1}}F^{bc}\d A_c\,,
\end{aligned}\ea
and
\ba\begin{aligned}
P^{abcd}=\frac{1}{2}\lf(g^{ac}g^{bd}-g^{ad}g^{bc}\rt)+\a_\text{GB} \frac{\pd \math{L}_\text{GB}}{\pd R_{abcd}}\,.
\end{aligned}\ea
The symplectic current $(n-1)$-form is defined by
\ba\begin{aligned}
\w(\f,\d_1\f,\d_2\f)=\d_1\bm{\Q}(\f,\d_2\f)-\d_2\bm{\Q}(\f,\d_1\f)\,.
\end{aligned}\ea
Utilizing \eq{Q2}, one can further obtain
\ba\begin{aligned}
\bm{\w}(\f,\d_1\f,\d_2\f)=\bm{\w}^\text{GB}(\f,\d_1\f,\d_2\f)+\bm{\w}^\text{EM}(\f,\d_1\f,\d_2\f)
\end{aligned}\nn\\\ea
with
\ba\begin{aligned}\label{w2}
&\bm{\w}_{a_1\cdots a_{n-1}}^\text{GB}(\f,\d_1\f,\d_2\f)=\frac{1}{8\p}\lf[\d_1\lf(\bm{\epsilon}_{ba_1\cdots a_{n-1}}P_a{}^{cbd}\rt)\d_2 \G^a{}_{cd}\right.\\
&\left.-\d_2\lf(\bm{\epsilon}_{ba_1\cdots a_{n-1}}P_a{}^{cbd}\rt)\d_1 \G^a{}_{cd}+\d_1(\bm{\epsilon}_{ba_1\cdots a_{n-1}}\grad_a P^{acbd})\d_2 g_{bd}\right.\\
&+\left.-\d_1(\bm{\epsilon}_{ba_2\cdots a_{n-1}}\grad_a P^{acbd})\d_1 g_{bd}\right]\\
&\bm{\w}_{a_1\cdots a_{n-1}}^\text{EM}(\f,\d_1\f,\d_2\f)=-\frac{1}{4\p}\left[\d_1(\bm{\epsilon}_{ba_1\cdots a_{n-1}}F^{bc})\d_2 A_c\right.\\
&\left.-\d_2(\bm{\epsilon}_{ba_1\cdots a_{n-1}}F^{bc})\d_1 A_c\right]\,.
\end{aligned}\ea
Replacing the variation by a infinitesimal diffeomorphism generated by the vector field $\z^a$, according to Eqs. \eq{action} and \eq{var1}, we can define a Noether current $(n-1)$-form as
\ba\begin{aligned}\label{J1}
\bm{J}_\z=\bm{\Q}(\f,\math{L}_\z\f)-\z\.\bm{L}\,.
\end{aligned}\ea
It has been shown in \cite{Wald94} that it can be expressed as
\ba\begin{aligned}\label{J2}
\bm{J}_\z=\bm{C}_\z+d\bm{Q}_\z\,,
\end{aligned}\ea
in which $C_\z=\z\.\bm{C}$ with
\ba\begin{aligned}
\bm{C}_{aa_2\cdots a_{n-1}}&=\bm{\epsilon}_{ba_2\cdots a_{n-1}}(T_a{}^b+A_aj^b)
\end{aligned}\ea
is the constraint of the Gauss-Bonnet gravity, and
\ba\begin{aligned}
\bm{Q}_\z&=\bm{Q}_\z^\text{GB}+\bm{Q}_\z^\text{EM}\\
\end{aligned}\ea
with
\ba\begin{aligned}\label{Q22}
(\bm{Q}_\z^\text{GB})_{a_1\cdots a_{n-2}}&=-\frac{1}{16\p}\bm{\epsilon}_{aba_1\cdots a_{n-2}}\\
&\times\left(P^{abcd}\grad_c\x_d-2\x_d\grad_cP^{abcd}\right)\,,\\
(\bm{Q}_\z^\text{EM})_{a_1\cdots a_{n-2}}&=-\frac{1}{8\p}\bm{\epsilon}_{aba_1\cdots a_{n-2}}F^{ab}A_c\z^c
\end{aligned}\ea
is the $(n-2)$-form Noether charge. Replacing $\z^a$ by the Killing vector $\x^a=(\pd/\pd v)^a$ of the background geometry in Eqs. \eq{J1} and \eq{J2}, the first two order variational identities can be further obtained,
\ba\begin{aligned}
d[\d\bm{Q}_\x-\x\.\bm{\Q}(\f,\d\f)]&+\d \bm{C}_\x=0\,,\\
d[\d^2\bm{Q}_\x-\x\.\d\bm{\Q}(\f,\d\f)]&=\bm{\w}\lf(\f,\d\f,\math{L}_\x\d\f\rt)\\
&-\x\.\d\bm{E}_\f\d\f-\d^2 \bm{C}_\x\,.
\end{aligned}\ea
Here we have used the fact that $T_{ab}=j^b=0$ for the background fields and $\x^a$ is a Killing vector so that $\math{L}_\x\f=0$. Integrating the first variational identity on the hypersurface $\S$ as introduced in the last section, we have
\ba\begin{aligned}\label{var11}
\int_{S_\inf}\lf[\d\bm{Q}_\x-\x\.\bm{\Q}(\f,\d\f)\rt]+\int_{\S_0}\d \bm{C}_\x+\int_{\S_1}\d \bm{C}_\x=0\,,
\end{aligned}\ea
where we used the assumption that the perturbation vanishes on the bifurcation surface $B$. For the first term, the integration is on the infinity boundary of $\S_1$, therefore, here the fields can be described by \eq{dsAS}. Then, the gravity part can be straightly calculated and it gives
\ba\begin{aligned}\label{dM1}
\int_{S_\inf}\lf[\d\bm{Q}_\x^\text{GB}-\x\.\bm{\Q}^\text{GB}(\f,\d\f)\rt]=\d M\,.
\end{aligned}\ea
For the EM part, using \eq{dsAS}, we have
\ba\begin{aligned}\label{QEM}
\bm{Q}_\x^\text{EM}(\l)=-\frac{4\p Q^2(\l)}{\W^2(D-3)r^{D-2}}\lf(\frac{1}{r_h^{D-3}}-\frac{1}{r^{D-3}}\rt)\hat{\bm{\epsilon}}\,,
\end{aligned}\ea
on $S_\inf$, where $\hat{\bm{\epsilon}}$ is the volume element of the $(n-2)$-dimensional sphere, which can be expressed as
\ba\begin{aligned}
\hat{\bm{\epsilon}}(\l)=r^{n-2}\left[\prod_{i=1}^{n-3}\sin^{n-2-i}\q_i\right]d\q_1\wedge \cdots \wedge d\q_{n-2}\,.
\end{aligned}\ea
The above expression then gives
\ba\begin{aligned}
\d \bm{Q}_\x^\text{EM}&=\left.\frac{d \bm{Q}_\x^\text{EM}}{d\l}\right|_{\l=0}\\
&=-\frac{8\p Q \d Q}{\W^2(D-3)r^{D-2}}\lf(\frac{1}{r_h^{D-3}}-\frac{1}{r^{D-3}}\rt)\hat{\bm{\epsilon}}\,.
\end{aligned}\ea
Integrating it, we can further obtain
\ba\begin{aligned}
\int_{S_\inf}\d \bm{Q}_\x^\text{EM}=-\frac{8\p Q \d Q}{\W(D-3)r^{D-3}_h}=-2\F_H\d Q\,.
\end{aligned}\ea
Using \eq{Q2}, we can also obtain
\ba\begin{aligned}
\int_{S_\inf}\x\.\bm{\Q}^\text{EM}(\f,\d\f)=-\F_H\d Q\,.
\end{aligned}\ea
Then, we have
\ba\begin{aligned}
\int_{S_\inf}\lf[\d\bm{Q}_\x^\text{EM}-\x\.\bm{\Q}^\text{EM}(\f,\d\f)\rt]=-\F_H\d Q\,.
\end{aligned}\ea

Using the stability assumption that $T_{ab}(\l)=j^a(\l)=0$ on $\S_1$, the third term of \eq{dM1} vanishes. Combining above results, we have
\ba\begin{aligned}\label{ineq1}
\d M-\F_H\d Q&=-\int_{\S_0}\d \bm{C}_\x\\
&=-\int_{\S_0}\d \lf[\bm{\epsilon}_{ba_2\cdots a_{D-1}}(\x^aT_a{}^b+\x^aA_aj^b)\rt]\\
&= \d\left[\int_{\S_0}\bm{\tilde{\epsilon}} T_{ab}(dr)^a\x^b\right]\,.
\end{aligned}\ea
The last step we used the gauge condition $\left.\x^aA_a(\l)\right|_{r=r_h}=0$. Here $\tilde{\bm{\epsilon}}$ is the volume element on $\S_0$, which is defined by
$\tilde{\bm{\epsilon}}= dv\wedge \hat{\bm{\epsilon}}\,.$ Next, we turn to calculate the second-order variational identity in \eq{var1}. With similar consideration and integrating it on $\S$, we have
\ba\begin{aligned}\label{eq2}
\int_{S_\inf}\left[\d^2 \bm{Q}_\x-\x\.\d\bm{\Q}(\f,\d\f)\right]&=\math{E}_{\S_0}-\int_{\S_0}\d^2\bm{C}_\x
\end{aligned}\ea
where we denote
\ba\begin{aligned}
\math{E}_0=\int_{\S_0}\bm{\w}(\f,\d\f,\math{L}_\x\d\f)\,.
\end{aligned}\ea
Here we used that $\math{L}_\x \f(\l)=T_{ab}(\l)=j^a(\l)=0$ on $\S_1$ from \eq{dsAS}, and $\x^a$ is a tangent vector on $\S_0$. With Straight calculation based on the explicit expressions on $\S_1$, it is not difficult to verify
$\math{E}_0=0$ in our case. For the  gravity part in left side in \eq{eq2}, similar calculation gives
\ba\begin{aligned}\label{dM2}
\int_{S_\inf}\left[\d^2 \bm{Q}_\x^\text{GB}-\x\.\d\bm{\Q}^\text{GB}(\f,\d\f)\right]=\d^2M\,.
\end{aligned}\ea
For the EM part, according to \eq{QEM}, we can further obtain
\ba\begin{aligned}
\d^2\bm{Q}_\x^\text{EM}=-\frac{8\p(Q\d^2Q+\d Q^2)}{\W^2(n-3)r^{n-2}}\lf(\frac{1}{r_h^{n-3}}-\frac{1}{r^{n-3}}\rt)\hat{\bm{\epsilon}}\,,
\end{aligned}\ea
which gives
\ba\begin{aligned}
\int_{S_\inf}\d^2 \bm{Q}_\x^\text{EM}&=-\frac{8\p (Q \d^2 Q+\d Q^2)}{\W(n-3)r^{n-3}_h}\\
&=-\frac{8\p\d Q^2}{\W(n-3)r^{n-3}_h}-2\F_H\d^2 Q\,.
\end{aligned}\ea
With similar calculation, we also have
\ba\begin{aligned}
\int_{S_\inf}\x\.\d\bm{\Q}^\text{EM}(\f,\d\f)=-\frac{4\p\d Q^2}{\W(n-3)r^{n-3}_h}-\F_H\d^2 Q\,.
\end{aligned}\ea
Then, the second-order variational identity reduces to
\ba\begin{aligned}
\d^2M-\F_H\d^2Q-\frac{4\p\d Q^2}{\W(n-3)r^{n-3}_h}&=-\int_{\S_0}\d^2\bm{C}_\x\\
&=\d^2\left[\int_{\S_0}\bm{\tilde{\epsilon}} T_{ab}(dr)^a\x^b\right]\,.
\end{aligned}\ea

Finally, we consider the null energy condition. Note that in the above setup, $\x^a$ is not a null vector on $\S_0$ for the geometry with the metric $g_{ab}(\l)$. Therefore, in order to obtain a relevant null energy condition, here we choose the null vector field as
\ba\begin{aligned}
l^a(\l)=\x^a+\b(\l)r^a\,.
\end{aligned}\ea
with
\ba\begin{aligned}
r^a=\lf(\frac{\pd}{\pd r}\rt)^a\,,\ \ \ \b(\l)=\frac{f^{(\l)}(r,v)}{2}\,.
\end{aligned}\ea
We can see that $\b=\b(0)=0$ on $\S_0$ since $f^{(0)}(r_h,v)=f(r_h)=0$. Then, we have
\ba\begin{aligned}
&T_{ab}(\l)l^a(\l)l^b(\l)\\
&=T_{ab}(\l)\x^a (dr)^b+\b^2(\l)T_{ab}(\l)r^ar^b\,.
\end{aligned}\ea
Considering the fact that $T_{ab}=0$ and $\b=0$ for the background geometry, we have
\ba\begin{aligned}
\d\left[\bm{\tilde{\epsilon}} T_{ab}(dr)^a\x^b\right]&= \d\left[T_{ab}l^al^bdv\wedge \hat{\bm{\epsilon}}\right]\,,\\
\d^2\left[\bm{\tilde{\epsilon}} T_{ab}(dr)^a\x^b\right]&= \d^2\left[T_{ab}l^al^bdv\wedge \hat{\bm{\epsilon}}\right]\,.
\end{aligned}\ea
Under the second-order approximation of perturbation, the null energy condition gives
\ba\begin{aligned}
&\int_{\S_0}T_{ab}(\l)l^a(\l)l^b(\l)dv\wedge \hat{\bm{\epsilon}}(\l)\\
&\simeq\l \d\left[\int_{\S_0}\bm{\tilde{\epsilon}} T_{ab}(dr)^a\x^b\right]+\frac{\l^2}{2}\d^2\left[\int_{\S_0}\bm{\tilde{\epsilon}} T_{ab}(dr)^a\x^b\right]\geq 0\,.
\end{aligned}\ea
Using the first-order and second-order variational identities, this inequality gives
\ba\begin{aligned}
\d M-\F_H\d Q+\frac{\l}{2}\left[\d^2M-\F_H\d^2Q-\frac{4\p\d Q^2}{\W(n-3)r^{n-3}_h}\right]\geq 0\,,
\end{aligned}\nn\\\ea
which can also be expressed as
\ba\begin{aligned}\label{inequality}
\d \m-\frac{2q\d q}{r^{n-3}_h}+\frac{\l}{2}\left[\d^2 \m-\frac{2q\d^2 q}{r^{n-3}_h}-\frac{2\d q^2}{r_h^{n-3}}\right]\geq0\,.
\end{aligned}\ea

\section{Gedanken experiments to overcharge the nearly extremal static charged Gauss-Bonnet black holes}\label{sec4}

In this section, we would like to consider the new version of the gedanken experiments to overcharge the static charged Gauss-Bonnet black holes. By virtue of the stability assumption, examining the weak cosmic censorship conjecture is equivalent to checking whether the geometry at sufficient late times describes a black hole solution, i.e., $f^{(\l)}(r)$ has at least one root. For simplification, here we can equivalently consider the function
\ba\begin{aligned}
p(r,\m,q)=\m-\left(q^2r^{n-3}+\frac{r^2+\a}{r^{n-5}}\right)\,.
\end{aligned}\ea
Its largest root will give the horizon of the black hole. Therefore, here we define a function
\ba\label{cd1}
h(\l)=p\lf(r_m(\l),\m(\l),q(\l)\rt)\,,
\ea
to describe the maximal value of $p(r, \m(\l), q(\l))$. Here $r_m(\l)$ is the maximal radius of the function $p(r, \m(\l), q(\l))$, and it satisfies
\ba\begin{aligned}\label{rmeq}
\pd_rp\lf(r_m(\l),\m(\l),q(\l)\rt)=0\,,
\end{aligned}\ea
which gives
\ba\begin{aligned}
q^2&=\frac{1}{r_m^{2n-8}}\left(r_m^2+\frac{(n-5)\a}{n-3}\right)\,,\\
\d r_m&=\frac{(n-3)r_m^{9-2n}q\d q}{(n-3)^2r_m^2+(n-5)(n-4)\a}\,.
\end{aligned}\ea

If the configurations $\f(\l)$ give $h(\l)>0$, the WCCC is violated. Under the second-order approximation of $\l$, we have
\ba\begin{aligned}\label{hl1}
&h(\l)=\m-\frac{2}{r_m^{D-5}}\left(r_m^2+\frac{(n-4)\a}{n-3}\right)+\l \left(\d M-\frac{2q\d q}{r_m^{n-3}}\right)\\
&+\frac{\l^2}{2}\lf(\d^2\m-\frac{2q\d^2q}{r_m^{n-3}}-\frac{2\d q^2}{r_m^{n-3}}\rt)-2\frac{\l^2(n-3)q}{r_m^{n-2}}\d r_m \d q\\
&-\l^2\left[(n-3)^2r_m^2+(n-4)(n-5)\a\right]\d r_m^2\\
\end{aligned}\ea

In the following, we consider the nearly extremal black hole cases where the maximal radius can be expressed as $r_m=(1-\epsilon)r_h$ with a small parameter $\epsilon$. Following a similar setup with \cite{SW}, we choose $\epsilon$ to agree with the first-order perturbation of the matter source. Under the first-order approximation of $\epsilon$, Eq. \eq{rmeq} gives
\ba
\pd_rp\lf(r_h,\m,q\rt)=\epsilon r_h \pd_r^2p\lf(r_h,\m,q\rt)\,.
\ea
Then, we have
\ba\begin{aligned}
&p(r_m,\m,q)=p\lf(r_h(1-\epsilon), \m, q\rt)\\
&\simeq -\epsilon r_h \pd_r p(r_h,\m,q)+\frac{\epsilon^2 r_h^2}{2} \pd_r^2 p(r_h,\m,q)\\
&=-\frac{1}{2}\epsilon^2 r_h^2 \pd_r^2 p(r_h,\m,q)
\end{aligned}\ea
at the second order of $\epsilon$, which gives
\ba\begin{aligned}
&\m-\frac{2}{r_m^{n-5}}\left(r_m^2+\frac{(n-4)\a}{n-3}\right)\\
&=\epsilon^2r_e^{n-5}\left[(n-3)^2r_e^2+(n-4)(n-5)\a\right]
\end{aligned}\ea

Combining the above results utilizing the perturbation inequality \eq{inequality}, Eq. \eq{hl1} becomes
\ba\begin{aligned}
&h(\l)=\epsilon^2r_e^{n-5}\left[(n-3)^2r_e^2+(n-4)(n-5)\a\right]\\
&-\frac{2(n-3)\l q \d q\epsilon}{r_e^{n-3}}+\frac{(n-3)^2q^2r_e^{11-3n}\d q^2}{(n-3)^2r_e^2+(n-4)(n-5)\a}\\
&=\frac{(n-3)\left[r^{2n}_e\epsilon+q r_e^6\left(q\epsilon(n-4)-\l \d q\right)\right]^2}{(n-4)q^2r_e^{n+9}+r_e^{3n+3}}
\end{aligned}\ea
under the second-order approximation of perturbation. In the last step, we have replaced $r_h$ by the horizon radius $r_e$ of the extremal case because we can neglect the difference of $r_m$ and $r_h$ at the second order in the above calculation. Then, we have $h(\l)>0$ under the second-order approximation of perturbation, which implies that the static charged Gauss-Bonnet black holes cannot be overcharged in the above collision process.

\section{conclusion}\label{sec5}

In this paper, following the setup of the gedanken experiments proposed by Sorce and Wald \cite{SW}, we tested the weak cosmic censorship conjecture in EMGB gravity under the perturbation of the spherically charged  infalling matter collision. First of all, based on the Iyer-Wald formalism, we derived the perturbation inequality under the second-order approximation when the matter fields satisfy the null energy condition and the collision satisfies the stability condition. Different with the result in \cite{SW}, here we did not utilize the optimal condition of the first-order identity. As a result, we found that the static charged Gauss-Bonnet black holes cannot be overcharged by above collision process under the second-order approximation of perturbation, although they can be destroyed by the old version of the gedanken experiments as shown in \cite{Ghosh:2019dzq}. Our result at some level implies that the WCCC can also be restored in the EMGB gravitational theory.

\section*{Acknowledgement}
This research was supported by National Natural Science Foundation of China (NSFC) with Grants No. 11775022 and 11873044.

\begin{equation*}
\end{equation*}


\begin{thebibliography}{100}
\bibitem{RPenrose}
 R. Penrose, Riv. Nuovo Cimento {\bf1}, 252 (1969).
\bibitem{Wald94}
R.M. Wald, Ann. Phys. (N.Y.) {\bf 82}, 548 (1974).
\bibitem{Hubeny}
V.E. Hubeny, Phys. Rev. D {\bf 59}, 064013 (1999).
\bibitem{B1}
T. Jacobson and T. P. Sotiriou, Phys. Rev. Lett. {\bf 103}, 141101 (2009).
\bibitem{B2}
T. Jacobson and T. P. Sotiriou, J. Phys. Conf. Ser. {\bf 222}, 012041 (2010).
\bibitem{B3}
G. E. A. Matsas and A. R. R. da Silva, Phys. Rev. Lett. {\bf 99}, 181301 (2007).
\bibitem{B4}
A. Saa and R. Santarelli, Phys. Rev. D {\bf 84}, 027501 (2011).
\bibitem{B5}
 S. Hod, Phys. Rev. D {\bf 66}, 024016 (2002).
\bibitem{B6}
S. Gao and Y. Zhang, Phys. Rev. D {\bf 87}, 044028 (2013).
\bibitem{B7}
Z. Li and C. Bambi, Phys. Rev. D {\bf 87}, 12, 124022 (2013).
\bibitem{B8}
K. Dzta and . Semiz, Phys. Rev. D {\bf 88}, 064043 (2013).
\bibitem{B9}
K. Dzta, Gen. Rel. Grav. {\bf 46}, 1709 (2014).
\bibitem{B10}
 G. Z. Tth, Class. Quant. Grav. {\bf 33}, 115012 (2016).
\bibitem{B11}
B. Gwak and B. H. Lee, JCAP {\bf1602}, 015 (2016).
\bibitem{B12}
B. Gwak and B. H. Lee, Phys. Lett. B {\bf 755}, 324 (2016).
\bibitem{B13}
V. Cardoso and L. Queimada, Gen. Rel. Grav. {\bf 47}, 12150 (2015).
\bibitem{B14}
 K. S. Revelar and I. Vega, Phys. Rev. D {\bf 96}, 064010 (2017).
\bibitem{B15}
J. Sorce and R. M. Wald, Phys. Rev. D {\bf 96}, 104014 (2017).
\bibitem{B16}
G. Chirco, S. Liberati and T. P. Sotiriou, Phys. Rev. D {\bf 82}, 104015 (2010).
\bibitem{B17}
H. M. Siahaan, Phys. Rev. D {\bf 93}, 064028 (2016).
\bibitem{SW}
J. Sorce and R.M. Wald, Phys. Rev. D {\bf 96}, 104014 (2017).
\bibitem{IW}
V. Iyer and R.M. Wald, Phys. Rev. D {\bf 50}, 846(1994).
\bibitem{An:2017phb}
  J.~An, J.~Shan, H.~Zhang and S.~Zhao, ``Five-dimensional Myers-Perry black holes cannot be overspun in gedanken experiments,''
  Phys.\ Rev.\ D {\bf 97}, 104007 (2018).
\bibitem{Ge:2017vun}
  B.~Ge, Y.~Mo, S.~Zhao and J.~Zheng, ``Higher-dimensional charged black holes cannot be over-charged by gedanken experiments,''
  Phys.\ Lett.\ B {\bf 783}, 440 (2018).
\bibitem{Jiang:2019ige}
  J.~Jiang, B.~Deng and Z.~Chen, ``Static charged dilaton black hole cannot be overcharged by gedanken experiments,''
  Phys.\ Rev.\ D {\bf 100}, 066024 (2019).
\bibitem{WJ}
  X.~Y.~Wang and J.~Jiang, ``Examining the weak cosmic censorship conjecture of RN-AdS black holes via the new version of the gedanken experiment,'' arXiv:1911.03938.
\bibitem{Jiang:2019vww}
  J.~Jiang, X.~Liu and M.~Zhang, ``Examining the weak cosmic censorship conjecture by gedanken experiments for Kerr-Sen black holes,'' Phys.\ Rev.\ D {\bf 100}, 084059 (2019).
\bibitem{Jiang:2020btc}
  J.~Jiang and M.~Zhang, ``Weak cosmic censorship conjecture in Einstein-Maxwell gravity with scalar hair,''
  Eur.\ Phys.\ J.\ C {\bf 80}, 196 (2020).
\bibitem{DWitten}
D.J. Gross and E. Witten, Nucl. Phys. {\bf B 277}, 1 (1986).
\bibitem{Zumino}
 B. Zumino, Phys. Rep. {\bf 137}, 109 (1986).
\bibitem{Ghosh:2019dzq}
  R.~Ghosh, C.~Fairoos and S.~Sarkar, ``Overcharging higher curvature black holes,'' Phys.\ Rev.\ D {\bf 100}, 124019 (2019).
\bibitem{GB1}
D. G. Boulware and S. Deser, String-Generated Gravity Models, Phys. Rev. Lett. {\bf55}, 2656 (1985).
\bibitem{GB2}
D. L. Wiltshire, Spherically symmetric solutions of Einstein-Maxwell theory with a Gauss-Bonnet term, Phys. Lett. {\bf 169B}, 36 (1986).
\bibitem{GB3}
 D. L. Wiltshire, Black holes in string-generated gravity models, Phys. Rev. D {\bf 38}, 2445 (1988).

\end{thebibliography}
\end{document}